\documentclass[pre,twocolumn,showpacs,preprintnumbers,amsmath,amssymb]{revtex4}
\usepackage{graphics,bm}

\begin{document}

\title{Heavy-tailed random error in quantum Monte Carlo}

\author{J. R. Trail} \email{jrt32@cam.ac.uk}
\affiliation{TCM Group, Cavendish Laboratory, University of Cambridge,
JJ Thomson Avenue, Cambridge, CB3 0HE, UK}

\date{May, 2007}

\begin{abstract}
The combination of continuum Many-Body Quantum physics and Monte Carlo methods 
provide a powerful and well established approach to first principles 
calculations for large systems.
Replacing the exact solution of the problem with a statistical estimate 
requires a measure of the random error in the estimate for it to be useful.
Such a measure of confidence is usually provided by assuming the Central Limit 
Theorem to hold true.
In what follows it is demonstrated that, for the most popular implementation 
of the Variational Monte Carlo method, the Central Limit Theorem has limited 
validity, or is invalid and must be replaced by a Generalised Central Limit 
Theorem.
Estimates of the total energy and the variance of the local energy are 
examined in detail, and shown to exhibit uncontrolled statistical errors 
through an 
explicit derivation of the distribution of the random error.
Several examples are given of estimated quantities for which the Central Limit 
Theorem is not valid.
The approach used is generally applicable to characterising the random error of
estimates, and to Quantum Monte Carlo methods beyond Variational Monte Carlo.
\end{abstract}

\pacs{02.70.Ss, 02.70.Tt, 31.25.-v}


\maketitle

Quantum Monte Carlo (QMC) provides a means of integrating over the full 
$3N$-dimensional coordinate space of a many-body quantum system in a 
computationally tractable manner while introducing a random error in the 
result of the integration\cite{foulkes01}.
The character of this random error is of primary importance to the 
applicability of QMC, and in what follows an understanding of the underlying 
statistics is sought for the special case of Variational Monte Carlo (VMC).

Within QMC, estimated expectation values have a random distribution of 
possible values, hence it is necessary to know the properties of this 
distribution in order to be satisfied that the statistical error is 
sufficiently well controlled.
Many strategies (notably those involved in wavefunction optimisation and total 
energy estimation) sample quantities that exhibit singularities, and sample 
the singularities rarely.
This is characteristic of a Monte Carlo (MC) strategy that is unstable and 
prone to abnormal statistical error due to outliers\cite{traub98}.

In what follows the VMC method is analysed in order to obtain the statistical 
properties of the random error.
Analytic results are obtained, and compared with the results of numerical 
calculations for an isolated all-electron carbon atom.
The analysis naturally divides into four sections.
Section \ref{sec:genvmc} provides a summary of the implementation of MC used 
within VMC.
The construction of estimated expectation value of an operator/trial 
wavefunction combination is described for the `standard sampling' case (the 
most commonly used form\cite{foulkes01}) as a special case of a more general 
formulation.
This short section provides no new results, but introduces the notation used 
throughout, and presents well established results from a perspective 
appropriate to the following sections.

Section \ref{sec:elsing} provides a transformation of the $3N$-dimensional 
statistical problem to an equivalent $1$-dimensional problem.
The purpose of this section is to provide a simple mathematical picture of the 
statistical process that is entirely equivalent to the original 
$3N$-dimensional random sampling process.
This is achieved by removing the statistical freedom in the system that is 
redundant for a given estimate.
The principal result of this section is the derivation of a general 
statistical property that arises for almost all of the trial wavefunctions 
available for VMC calculations, and that may not easily be prevented.
This statistical property dominates the behaviour of errors in VMC estimates, 
and the demonstration of its presence provides the starting point for the 
derivation of the statistics of estimators.

In section \ref{sec:std} the `standard sampling' formulation of VMC is 
analysed.
The goal is to find the distribution of the random error in statistical 
estimates of the total energy and the `variance', for a finite but large 
number of samples.
The principal conclusion of this section is that the Central Limit Theorem 
(CLT) is not necessarily valid and, when it is valid, finite sampling effects 
may be important even for a large sample size.
This is demonstrated analytically, in the form of new expressions for the 
distribution of errors occurring for `standard sampling' estimates of the 
total energy and variance.
Numerical results for an isolated carbon atom provide an example of this 
effect for a calculation employing an accurate trial wavefunction.

In section \ref{sec:otherestimates} estimates of several other quantities 
relevant to QMC are considered, and the invalidity of the CLT for these 
estimates is described (when derived using the same method as section 
\ref{sec:std}).
This section directly relates to the infinite variance estimators that 
have previously been discussed in the literature\cite{assaraf}.

Finally, we note that this is the first of two closely related papers.
It provides a general approach to rigorously deriving the statistics of the 
random error that is an inherent part of QMC methods, and uses this approach 
to obtain the statistical limitations of the simplest available sampling 
strategy.
The following paper\cite{trail07b} employs this new analysis of the statistics 
of QMC in order to design sampling strategies that are superior, in the sense 
that the Normal distribution of random errors can be reinstated for a given 
QMC estimate.

\section{`Standard sampling' Variational Monte Carlo}
\label{sec:genvmc}
The basic equation by which MC methods provide a statistical estimate for an 
integral may be written as 
\begin{equation}
\frac{1}{r} \sum_{n=1}^{r} \frac{ f(\text{\sffamily\bfseries R}_n) }
                                { P(\text{\sffamily\bfseries R}_n) } =
                           \int_V f d{\bf R} + \mathsf{W}_r
\label{eq:1}
\end{equation}
where $P$ is the probability density function (PDF) of the independent 
identically distributed (IID) $3N$-dimensional random vector 
$\text{\sffamily\bfseries R}_n$, and $\mathsf{W}_r$ is the random error in the 
estimate.

Introducing some notation used throughout the paper, the statistical 
estimate of a quantity $f$ constructed using $r$ samples is denoted 
$\mathsf{A}_r[f]$, hence Eq.~(\ref{eq:1}) can be written as
\begin{equation}
\mathsf{A}_r \left[ \int_V f d{\bf R} \right] = 
   \mathbb{E}\left[ \frac{f}{P};P \right] + \mathsf{W}_r,
\end{equation}
where the LHS is the statistical estimate of the integral (the sample 
mean in Eq.~(\ref{eq:1})), and the RHS can be interpreted as a sum of an 
expectation of a quantity $x=f/P$ sampled over the distribution with PDF 
$P$, and a random error.
Whether the estimate is useful depends on the PDF of $\mathsf{W}_r$, 
specifically how this distribution evolves as $r$ increases.

An expectation value of the quantum mechanical operator $\hat{g}$ and 
(unnormalised) wavefunction, $\psi$, is defined by
\begin{equation}
G = \frac{ \mathbb{E}\left[ G_L\psi^2/P;P \right] }
         { \mathbb{E}\left[    \psi^2/P;P \right] },
\label{eq:2}
\end{equation}
where $G_L=\psi^{-1}\hat{g} \psi $ is the `local value' of the operator/trial 
wavefunction combination.
By definition, VMC provides a MC estimate for this quantity, and since it is a 
quotient of two expectations it is more complex to estimate than a single 
integral.

`Standard sampling' is the most common and straightforward choice, for which 
samples are distributed as $P({\bf R})=\lambda \psi^2$, resulting in the 
simple form
\begin{eqnarray}
\mathsf{A}_r\left[G\right] &=& 
    \mathbb{E}\left[ G_L ;\lambda \psi^2\right] + \mathsf{Y}_r \nonumber \\
                             &=& \frac{1}{r}\sum_{n=1}^{r} 
    G_L( \text{\sffamily\bfseries R}_n ) , \;\;\;\; P({\bf R})=\lambda \psi^2,
\end{eqnarray}
where $\lambda$ need not be known since it is not required to generate samples 
distributed as $P({\bf R})$\cite{foulkes01}.
This simple form arises from choosing $P$ such that the normalisation integral 
of Eq.~(\ref{eq:2}) is sampled perfectly.

Within `standard sampling' it is usually \textit{assumed} that the CLT is 
valid, and that $r$ is large enough for the asymptotic limit to be reached to 
a required accuracy.
If this is so, then $\mathsf{Y}_r$ is distributed normally with a mean of $0$, 
a variance given in terms of the sample variance
\begin{equation}
\mathrm{Var} \left[ \mathsf{A}_r\left[ G \right]\right] = 
\frac{1}{r}\;
\mathsf{A}_r \left[ \mathrm{Var} \left[ 
G_L( \text{\sffamily\bfseries R}_n )
\right] \right] ,
\end{equation}
and a confidence range for an estimated value can be obtained via the error 
function.

Two issues concerning the nature of the random error naturally suggest 
themselves.
The use of the CLT to provide a confidence interval for the estimate 
implicitly assumes that the large $r$ limit has been reached.
Whether this is the case for finite $r$ is a non-trivial 
question\cite{stroock93}.
The second issue is the validity of the CLT.
Since this theorem is applicable to a limited class of distributions that may 
or may not include the distribution of samples within VMC 
(or other QMC methods) this is also a non-trivial question.

It is useful at this point to introduce some further definitions and notation.
An estimate is a random variable, and random variables are denoted by a 
sans-serif font throughout.
A particular sample value of an estimate is referred to as a sample estimate, 
and estimates are usually constructed from sums of random variables.
The PDF of the estimate constructed from $r$ random variables is denoted 
$P_r(x)$, and defined by
\begin{equation}
\textrm{Prob}\left[ a < \mathsf{A}_r\left[ G \right] \leq b \right] = 
  \int_a^b P_r(x) dx,
\end{equation}
and an estimate is unbiased if it has a mean for a given $r$ that is equal to 
its true value.
For the estimate to be useful the PDF of the error, $\mathsf{Y}_r$, must 
possess certain properties.
It would be desirable for this PDF to approach a Dirac delta function for 
increasing $r$, and for some information to be available on the form of the 
PDF for finite $r$.
In addition an estimate-able confidence range for finite $r$ is desirable, and 
zero mean value for $\mathsf{Y}_r$ for finite $r$.

\section{General asymptotic form for the distribution of local energies}
\label{sec:elsing}
For the standard implementation of VMC summarised in the previous section, the 
basic random variable is the $3N$-dimensional position vector of all the 
particles within the system, $\text{\sffamily\bfseries R}$.
This is a `fundamental' random variable in the sense that QMC is normally 
implemented as a random walk in the multidimensional space, $\mathbf{R}$.
However, this random variable contains far more information than is required 
for many purposes.
An analysis is given here for the expectation value of quantities that may be 
expressed in terms of the local energy, $E_L=\psi^{-1} \hat{H}\psi$.
Note that this is a general procedure, and is applicable to estimates of any 
operator by defining a local field variable (scaler, vector or higher order) 
to remove the redundant statistical freedom present in the full 
$3N$-dimensional space, providing a more concise representation.

The expectation of a function of the local energy $E_L$ is defined as
\begin{eqnarray}
\frac{\langle \psi | f(\hat{H}) | \psi \rangle}{\langle \psi | \psi \rangle}
         & = & \mathbb{E}\left[ f;\lambda \psi^2 \right]  \\
         & = & \int P_{\psi^2}( \mathbf{R} ) f( E_L ) d\mathbf{R},
\end{eqnarray}
and the `standard sampling' MC estimate of this is constructed by sampling the 
$3N$-dimensional coordinate vector over the `seed' PDF 
$P_{\psi^2}( \mathbf{R} )=\lambda \psi^2$.

Integrating over a hyper-surface of constant local energy removes redundant 
statistical degrees of freedom leaving the field variable, $E_L(\mathbf{R})$, 
as the random variable.
The expectation is then given by
\begin{equation}
\mathbb{E}\left[ f;P_{\psi^2} \right] 
   = \int P_{\psi^2}( E ) f( E ) dE,
\end{equation}
with the `seed' PDF of the local energy given by
\begin{equation}
P_{\psi^2}( E ) 
   = \int_{\partial} 
     \frac{P(\mathbf{R})}
          {\left| \nabla_{\mathbf{R}} E_L \right|} d^{3N-1}\mathbf{R},
\label{eq:3}
\end{equation}
where ${\partial}$ is a surface of constant $E_L$, and 
$\nabla_{\mathbf{R}} E_L$ is the gradient of the local energy in 
$3N$-dimensional space.
The interpretation of this surface integral is straightforward, provided that 
disconnected surfaces and non-smoothness in the hyper-surface are dealt with 
as a sum of separate (and sometimes connected) surface integrals.
Equation~(\ref{eq:3}) simplifies the interpretation of general statistical 
properties considerably.
Analytic properties of the seed distribution may be derived that are general 
to the $(\hat{H},\psi)$ combinations used for VMC.

In what follows we limit ourselves to the case of electrons in the potential 
of fixed atomic nuclei and Coulomb interactions, giving a local energy in 
$3N$-dimensional space of the form
\begin{eqnarray}
E_L(\mathbf{R}) &=& -\frac{1}{2}\frac{\nabla^2_{\mathbf{R}} \psi}{\psi} 
                    + V_{ee}(\mathbf{R}) + V_{ext}(\mathbf{R }) \nonumber \\
                &=& T_L + V_L,
\end{eqnarray}
where $V_{ee}(\mathbf{R})$ is the sum of two-body potentials (the 
electron-electron Coulomb interaction), and $V_{ext}(\mathbf{R})$ is the sum 
of one body potentials (the electron-nucleus Coulomb interaction).
$T_L$ is the local kinetic energy, and all other terms are contained in $V_L$, 
the local potential energy.
Singularities will occur for a general $\psi$, and the expression above 
naturally suggests classifying these into  $4$ different types.
Each has a characteristic influence on the asymptotic behaviour of 
$P_{\psi^2}( E )$, and an analysis of this relationship is given below.

\begin{figure*}[t]
\includegraphics{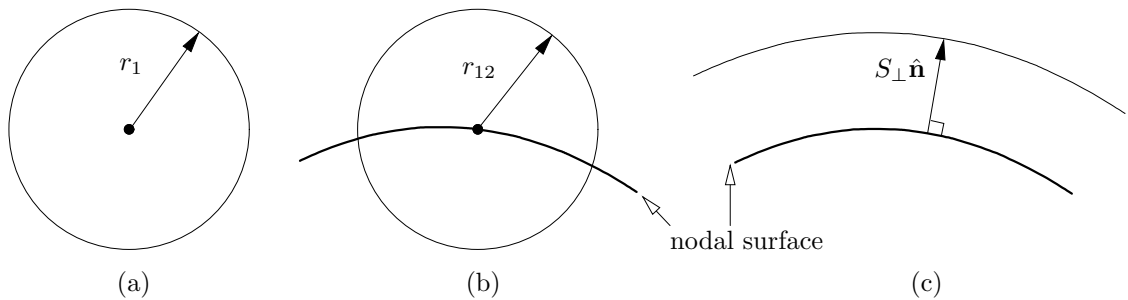}
\caption{\label{fig1}  Constant energy surfaces in the large $E$ limit.
Figure (a) shows the surface in terms of the electron-nucleus vector as 
coalescence is approached. The same geometry arises for electron-electron 
coalescence where the electrons possess different spin.
Figure (b) shows the surface in terms of the electron-electron vector for 
electrons of like spin for the case where no singularity is present in the 
local kinetic energy.
Figure (c) shows the constant energy surface for singularities at the nodal 
surface due to the local kinetic energy, $T_L$.
}
\end{figure*}

\subsection{Type 1: electron-nucleus coalescence}
Type 1 singularities are those resulting from any electron coordinate 
$\mathbf{r}_i$ approaching a singularity in the one body external potential 
$V_{ext}$, such as the $-Z/r$ behaviour of an atomic nucleus.
This occurs on a $3N-3$ dimensional hyper-surface.

For a particular electron of coordinate $\mathbf{r}_1$ approaching a nucleus, 
the trial wavefunction can be expanded in spherical coordinates to give
\begin{equation}
\psi(\mathbf{r}_1) = a_0( \mathbf{R}_{3N-3} ) + 
   a_1( \Omega, \mathbf{R}_{3N-3} )r_1 + \ldots,
\end{equation}
where $\mathbf{r}_1 = ( r_1,\Omega )$ and $\mathbf{R}_{3N-3}$ is the ${3N-3}$ 
dimensional vector of the rest of the coordinate space.
If $\psi$ does not possess singularities, it must be possible to expand 
$a_n(\Omega)$ as a closed sum of spherical harmonics $Y_{lm}(\Omega)$ with 
$l \leq n$.
Similarly, for $\psi$ to be continuous up to order $n$, the coefficient 
$a_n(\Omega)$ must contain only odd/even $l$ spherical harmonics in its 
expansion for odd/even $n$.

For a trial wavefunction that is smooth at $r_1=0$ this results in a local 
energy of the form
\begin{equation}
E_L(\mathbf{R}) - E_0 =-\frac{Z}{r_1} + b_{ 0}( \mathbf{R}_{3N-3}) +  \ldots.
\end{equation}
The absence of a $r_1^{-2}$ term is a direct consequence of $\psi$ being 
continuous at $r_1=0$, and the $r_1^{-1}$ term is entirely due to the presence 
of the nucleus potential and the derivative of $\psi$ being continuous at 
$r_1=0$.
Figure~\ref{fig1}(a) shows a 2D cut through the 3D space of $\mathbf{r}_1$, 
with $\mathbf{R}_{3N-3}$ held constant and the singularity due to the nucleus 
at the centre of the (asymptotically) spherical constant energy surface.

Rearranging and repeated re-substitution provides the integrand in 
Eq.~(\ref{eq:3}), and integrating over the constant energy surface defined by 
the limit $E_{L} \rightarrow \pm \infty$ (a `hyper-tube' which is spherical in 
the space of $\mathbf{r}_1$, but has no simple form in the $3N-3$ dimensions 
of $\mathbf{R}_{3N-3}$) gives the general form for the tail
\footnote{To put this more explicitly, the integrand in Eq.~(\ref{eq:3}) 
is expressed as a ratio of two power series in $r_1$, then re-expanded as a 
single power series in $r_1$ (series are only required to converge for 
$r_1$ close to $0$). A general Jacobian is included.
Next it is noted that as $E_L \rightarrow \pm \infty$ the constant energy 
surface approaches a sphere in the sub-space $\mathbf{r}_1$, so the $0^{th}$ 
and $1^{st}$ order dependence of the Jacobian on $r_1$ approaches zero.
The surface integral then results in a function of energy only (the energy 
of the constant energy surface).
In essence this provides the asymptotic form of $P_{\psi^2}$ resulting
from the chosen form of wavefunction and Hamiltonian, and no integrals 
are required explicitly.}
\begin{equation}
P_{\psi^2}(E) =
  \left\{
  \begin{array}{ll}
    0                                                         & E \gg E_0 \\
    (E-E_0)^{-4} \left(e_0+\frac{e_1}{(E-E_0)}+\ldots\right)  & E \ll E_0
  \end{array}
  \right. ,
\label{eq:3.1}
\end{equation}
where $E \gg E_0$ ($E \ll E_0$) denotes an asymptotic expansion that converges 
for $E$ greater (less) than some finite value.
The asymptotic behaviour is one sided since the singularity is negative, and 
the nodal surface does not need to be considered.

If the usual electron-nucleus Kato cusp condition\cite{pack66,myers91} is 
forced on $\psi$ it introduces a discontinuity in the gradient at $r_1=0$ that 
exactly cancels the singular nucleus potential in the local energy via the 
local kinetic energy, hence this type of singularity can generally be removed.
The cusp condition also introduces an $\Omega$ dependence in the $b_0$ term of 
the expansion, and hence a discontinuity in the local energy at the nucleus 
(although it is of zero size for some wavefunctions)
\footnote{
Taking a general smooth wavefunction and applying an appropriate
cusp correction results in a new wavefunction, $\psi$, that satisfies the
Kato cusp condition.  This may be expanded as a power series in the
electron-nucleus vector $\mathbf{r}_1$ :
$\psi(\mathbf{r}_1)=
  a + \mathbf{b}.\mathbf{r}_1 - a Z r_1 +\mathcal{O}(r_1^2 )$.
It is straightforward to show that $E_L=\psi^{-1} \hat{H}\psi$
possesses no singularity at $\mathbf{r}_1=0$, but is discontinuous
unless $\mathbf{b}=0$.  There are many examples of wavefunctions for
which $\mathbf{b}=0$, such as the exact wavefunction, or a Slater
determinant of exact Hartree-Fock orbitals, but this is not a
consequence of satisfying the Kato cusp condition.
Note that this analysis is only valid when $\psi$ is finite at the
nucleus.  If $\psi$ is zero at the nucleus, then the absence of a
singularity and continuity of the local energy at the nucleus require
two new conditions to be satisfied which replace the Kato cusp and
$\mathbf{b}=0$ conditions.  These may easily be derived.  Note that
this analysis does not imply any statement about the continuity of the
local energy as two or more electrons coalesce at a
nucleus\cite{myers91}.
}.
For $N$ electrons approaching the nucleus concurrently the same cusp 
conditions are sufficient to prevent a singularity, as discussed in the next 
section.

\subsection{Type 2: electron-electron coalescence}
Type 2 singularities may occur for $\mathbf{r}_i$ approaching $\mathbf{r}_j$ 
($i \ne j$), and result from a singularity in the two-body electron-electron 
interaction, $V_{ee}$.
The coalescence of electrons of like spin (indistinguishable) and unlike spin 
(distinguishable) must be considered separately.
By transforming to centre of mass coordinates for the two electrons with 
positions vectors $\mathbf{r}_{1},\mathbf{r}_{2}$ defined as 
$\mathbf{r}_{12}=\mathbf{r}_1-\mathbf{r}_2$ and 
$\mathbf{s}_{12}=(\mathbf{r}_1+\mathbf{r}_2)/2$ the same approach can be taken 
as for the electron-nucleus coalescence surfaces.
To simplify the notation the vector $\mathbf{s}_{12}$ is included with the 
coordinates of the rest of the electrons in the vector $\mathbf{R}_{3N-3}$.

For \textit{distinguishable} (unlike spin) electrons the situation in entirely 
analogous to the electron-nucleus case.
The electron-nucleus vector and interaction is replaced by the 
electron-electron vector and interaction to give
\begin{equation}
P_{\psi^2}(E) =
  \left\{
  \begin{array}{ll}
    (E-E_0)^{-4} \left(e_0+\frac{e_1}{(E-E_0)}+\ldots\right)  & E \gg E_0 \\
    0                                                         & E \ll E_0
  \end{array}
  \right. ,
\end{equation}
where $E_0$ and the coefficients, $e_n$, are distinct from those in 
Eq.~(\ref{eq:3.1}).
(In order to keep the notation simple the same symbols are used for distinct 
coefficients in all of the series expansions contained within this section.)
The asymptote is one sided due to the repulsive electron-electron interaction, 
and the nodal surface does not influence the result.
Enforcing the Kato cusp condition for unlike spins removes these tails and 
introduces a discontinuity in the local energy in precisely the same manner as 
for the electron-nucleus coalescence.

For \textit{indistinguishable} (like spin) electrons the situation is more 
complex.
Figure~\ref{fig1}(b) shows a 2D cut through the 3D space of $\mathbf{r}_{12}$, 
with $\mathbf{R}_{3N-3}$ held constant and a constant energy surface that is 
(asymptotically) spherical in the electron-electron coordinate.
The singularity due to electron-electron coalescence is at the centre of the 
sphere.
Unlike the distinguishable electron case the coalescence point must fall on 
the nodal surface, and the influence this has on $\psi$ must be taken into 
account.

Expanding a smooth antisymmetric trial wavefunction about the coalescence 
point (on the nodal surface) gives 
\begin{equation}
\psi(\mathbf{r}_{12}) = a_1( \Omega, \mathbf{R}_{3N-3} )r_{12} + 
    a_3( \Omega, \mathbf{R}_{3N-3} )r_{12}^3+\ldots,
\end{equation}
where interchange of electrons corresponds to inversion about $r_{12}=0$ so 
the coefficient $a_n$ contains only odd $l$ spherical harmonics and $l \leq n$.
This provides a quadratic lowest order variation in the probability density 
perpendicular to the nodal surface, which results in a local energy of the form
\begin{equation}
E_L(\mathbf{R}) - E_0 = \frac{1}{r_{12}} + 
    b_{ 1}( \Omega,\mathbf{R}_{3N-3})r_{12} +  \ldots,
\end{equation}
and an (asymptotically) spherical constant local energy surface centred at the 
coalescence point.
Note that the absence of a $r_{12}^{-2}$ term is a direct consequence of the 
\textit{gradient} of $\psi$ being continuous at $r_{12}=0$.
The $r_{12}^{-1}$ term is entirely due to the Coulomb potential, together with 
$\psi$ being odd on interchange of electrons and possessing a continuous 
second derivative at $r_{12}=0$.
Performing the `hyper-tube' integration then gives
\begin{equation}
P_{\psi^2}(E) =
  \left\{
  \begin{array}{ll}
    (E-E_0)^{-6} \left(e_0+\frac{e_1}{(E-E_0)}+\ldots\right)  & E \gg E_0 \\
    0                                                         & E \ll E_0
  \end{array}
  \right. ,
\end{equation}
where, since the singularity is positive, the asymptotic behaviour is one 
sided.

Enforcing the Kato cusp condition\cite{pack66,myers91} for like spins 
introduces a second order radial term, with coefficient $a_2=a_1/4$.
This provides a discontinuity in the $2^{nd}$ order derivatives of $\psi$ at 
$r_{12}=0$ that cancels the singular electron-electron interaction, and so 
removes the tails due to the like spin electron-electron coalescence.
A further consequence is a continuous local energy as the coalescence plane is 
crossed, with a discontinuity in the gradient of the local energy.

So far only electron-nucleus and electron-electron coalescence has been 
considered.
For the general case of many electron coalescence (some distinguishable, some 
not) at a nucleus site, or at any point in space, and a smooth trial function 
$\psi$, the local energy may be written in the form
\begin{equation}
E_L(\mathbf{R}) - E_0 = \sum_{i} \frac{Z}{r_i} + 
                        \sum_{i<j} \frac{1}{r_{ij}} + \ldots
\end{equation}
provided that the local kinetic energy is smooth.
As discussed by Pack\cite{pack66}, provided the trial wavefunction satisfies 
the cusp conditions for each electron-electron and electron-nucleus 
coalescence, then the Coulomb singularities will exactly cancel with 
singularities in the local kinetic energy.
These conditions are easily satisfied for trial wavefunctions that are a 
function of electron-nucleus, electron-electron and electron-electron-nucleus 
coordinates, but for higher order correlations internal coordinates must be 
considered explicitly.

Although the Kato cusp conditions remove the Coulomb singularities from the 
local energy, they do not prevent the occurrence of discontinuities on the 
same hyper-surface of electron-nucleus and electron-electron coalescence.
Further cusp conditions that remove these discontinuities may be obtained 
directly from the local energy expansions given above.

\subsection{Type 3: nodal surface}
The third type of singularity (and associated tails in the seed distribution) 
occurs for almost all of the trial wavefunctions used in QMC calculations, 
with the exception of some few electron systems.

Type 3 singularities are due to the kinetic energy only, and occur at the 
nodal surface due to the presence of $\psi$ in the denominator of the 
expression for the local kinetic energy.
There is no equivalent to the previous cusp conditions that can easily be 
enforced on $\psi$ to prevent these type 3 singularities occurring, and they 
are of a fundamentally different nature.

Proceeding in a similar manner to the previous two cases, the trial 
wavefunction is expanded about the singular surface, in this case the $3N-1$ 
dimensional nodal surface.
This expansion is then used to provide a constant local energy hyper-surface, 
over which an integral is performed to obtain the PDF in energy space.

Figure~\ref{fig1}(c) shows a 2D cut through the $3N$ dimensional space that 
includes the nodal surface, and a constant local energy surface at a 
perpendicular distance $S_{\perp}$ from the nodal surface.
Expressing the vector of a point on the constant energy surface as
\begin{equation}
\mathbf{R} = \mathbf{X} + S_{\perp} \hat{ \mathbf{n} },
\end{equation}
where $\mathbf{X}$ is a point on the nodal surface, and 
$\hat{ \mathbf{n} }(\mathbf{X}) = 
\widehat{ \nabla_{\mathbf{R}} E_L} \left|_{\mathbf{X}}\right.$ is the 
normalised gradient at $\mathbf{X}$, gives 
\begin{equation}
\psi(\mathbf{R}) = a_1(\mathbf{X})S_{\perp}+a_2(\mathbf{X})S_{\perp}^2+\ldots
\end{equation}
and
\begin{equation}
E_L(\mathbf{R}) - E_0 = b_{-1}(\mathbf{X})S_{\perp}^{-1} 
                      + b_{0}(\mathbf{X}) + b_{1}(\mathbf{X})S_{\perp} 
                                      + \ldots.
\end{equation}
Employing these in Eq.~(\ref{eq:3}) and integrating over the constant energy 
surface defined by the limit $E_{L} \rightarrow \pm \infty$ 
(the nodal surface) gives the general form
\begin{equation}
P_{\psi^2}(E) =
  \left.
  \begin{array}{ll}
    (E-E_0)^{-4} \left(e_0+\frac{e_1}{(E-E_0)}+\ldots\right)  & 
    \ \ \ \ |E| \gg E_0 
  \end{array}
  \right. .
\label{eq:4}
\end{equation}
Equation~(\ref{eq:4}) tells us that for a general trial wavefunction and 
Hamiltonian the resulting `seed' probability distribution in energy space has 
this asymptotic form for type 3 singularities.
This result is central to the rest of this paper.

A special case of this type of singularity arises for a trial wavefunction 
where a nodal pocket is at the critical point of appearing/disappearing, which 
may occur in the process of varying a parameterised trial wavefunction in the 
search for an optimum form.
This occurs where a solution of the equation $\psi=0$ disappears, or for a 
local maximum/minimum of $\psi$ crossing the nodal surface.
At this critical point $\psi=0$ defines a single point in $3N$ dimensional 
space, and the wavefunction may be expanded about this point using 
hyper-spherical coordinates $\mathbf{R}=(R,\mathbf{\Omega})$ 
(with $R$ the hyper-radius and $\mathbf{\Omega}$ the $3N-1$ hyper-angles) as
\begin{equation}
\psi(\mathbf{R}) = a_2(\mathbf{\Omega})R^2+a_3(\mathbf{\Omega})R^3+\ldots.
\end{equation}
The associated local energy then takes the form
\begin{equation}
E_L(R,\mathbf{\Omega}) - E_0 = b_{-2}(\mathbf{\Omega})R^{-2} + 
                               b_{-1}(\mathbf{\Omega})R^{-1} +
                            b_{0} (\mathbf{\Omega})       + \ldots
\end{equation}
with the singular behaviour arising via the local kinetic energy.
Following the same approach as for type 1 and type 2 singularities, but 
integrating over the surface of the hyper-sphere gives
\begin{equation}
P_{\psi^2}(E) = \frac{1}{\left|E-E_0\right|^{(3N+6)/2}} 
                \left(e_0+\frac{e_1}{(E-E_0)}+\ldots\right),
\end{equation}
an asymptotic tail in the PDF that is one sided since the constant energy 
surface exists only in the nodal pocket that is not being created/annihilated.
This gives a faster decay than $E^{-4}$ for $N \ge 1$, and nodal pockets can 
only occur in the ground state for $N \ge 2$.
Consequently, this effect is secondary to the $E^{-4}$ behaviour arising from 
nodal surfaces that are not being created/annihilated, and will only dominate 
if annihilation of the nodal pocket results in no nodal surfaces anywhere in 
space.
This can only occur if all fermions in the system are distinguishable.

\subsection{Type 4: arbitrary bound trial wavefunctions}
Singularities in the local energy may also occur if the local energy 
approaches infinity as any or all electrons approach an infinite distance 
from the nuclei or each other.
This type of singularity is referred to as type 4, and its source may be the 
local kinetic energy, the local potential energy, or both, and can only occur 
for systems that do not extend over all space.

For these finite systems a reasonable assumption about the general form of a 
trial wavefunction used in QMC is that it is a bound state of some `model' 
Hamiltonian (this encompasses the exact, HF, MCSCF, Kohn-Sham, Gaussian basis 
wavefunctions, and many others, with or without a Jastrow factor or backflow 
transformation).
Hence, for the types of wavefunction that are used in QMC calculations, the 
asymptotic behaviour can be written as
\begin{equation}
\psi(\mathbf{R}) \propto \left|\mathbf{R}\right|^{\alpha} 
    e^{ -\beta \left|\mathbf{R}\right|^\gamma },
\end{equation}
where the parameters $\alpha$, $\beta$, and $\gamma$ depend on the type of 
trial wavefunction.

Following the same approach as for type 1 and 2 singularities, the influence 
on the asymptotic tails of the seed distribution can be determined by 
integrating over the constant local energy surface.
This tells us that for $\gamma > 1$ (e.g., a Gaussian basis set) $P_{\psi^2}$ 
decays as an exponential function of a power of $E$, whereas for 
$0 < \gamma \leq 1$ ($\gamma=1$ is the correct asymptotic form) $P_{\psi^2}$ 
is zero outside of an energy interval (assuming that none of the other 3 
types of singularity are present).
The second case is preferable, but the former is not significant as it can 
only result in the presence of exponentially decaying tails in $P_{\psi^2}$.
In what follows type 4 singularities are irrelevant.
\newline

Type 3 tails occur for almost all many body trial wavefunctions, with some 
exceptions.
First, it is possible for there to be no nodal surface.
This does not occur for systems containing two or more indistinguishable 
fermions, and does occur if the trial wavefunction is a bosonic ground state.
Second, the nodal surfaces may be exactly known from symmetry considerations, 
as discussed by Bajdich \textit{et al.}\cite{bajdich05}.
A third exception arises from considering an effective Hamiltonian for which 
the trial wavefunction is an exact solution.
This has a potential defined by
\begin{equation}
V_{eff}= E_{eff} + \frac{1}{2} \frac{ \nabla^2_{\mathbf{R}} \psi }{ \psi },
\end{equation}
where $E_{eff}$ is arbitrary, but is usually chosen to be zero for a 
completely ionized system.
If $V_{eff}$ can be shown to possess no singularities at the nodal 
surface, then $\nabla^2_{\mathbf{R}} \psi=0$ at the nodal surface and type 3 
tails do not occur.
An example is the Slater determinant, as this is the exact solution for 
fermions in a one-body potential (with no two-body or higher interactions 
present in $V_{eff}$).
(Note that the available modifications of such `exact model' solutions, 
such as Jastrow factors, result in a many-body $V_{eff}$ that is singular 
at the nodal surface.)

Removing type 3 singularities is a non-trivial problem since it is necessary 
to ensure that $T_L$ remains finite over the nodal surface apart from on the 
coalescence planes, where it must possess a singularity that exactly cancels 
the electron-electron Coulomb interaction.
Type 3 tails are taken to be unavoidable in practice.

In order to clarify when these singularities/tails occur it is  worth 
considering some examples.
For an exact wavefunction none of the singularity types occur.
For a Hartree-Fock or Kohn-Sham Slater determinant with no basis set error 
only type 2 singularities occur, since the electron-nucleus cusp conditions 
are satisfied, the asymptotic wavefunction behaviour has the correct 
exponential form, and the local kinetic energy is finite at the nodal surface.
For a Hartree-Fock or Kohn-Sham Slater determinant with a Gaussian basis set, 
singularities of all four types occur, but type 1 and 2 singularities can be 
expected to dominate.

\begin{figure}[t]
\includegraphics{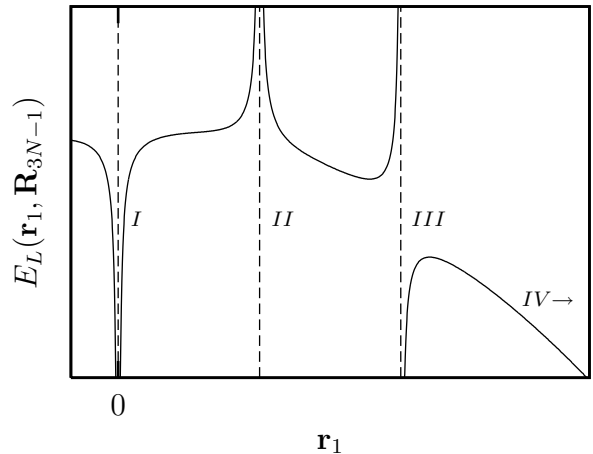}
\caption{\label{fig2}
Variation of the local energy in the presence of singularities of all four 
types, with an electronic coordinate, ${\bf r}_1$, passing through singular 
hyper-surfaces.
$I$, $II$, $III$, and $IV$ denote singularities due to e-n interaction, e-e 
interaction, the nodal surface, and incorrect asymptotic behaviour (shown here 
for the Gaussian case), respectively. Units are arbitrary.
}
\end{figure}

Figure~\ref{fig2} shows a schematic of the form taken by the singularities in 
the local energy as an electron passes through the nucleus, through a 
coalescence plane, through a nodal surface, and continues away from the 
nucleus, for the case where all types of singularity are present.
From this point on, only the influence of type 3 singularities and the 
associated symmetric tails in the seed distribution are considered, since 
type 1 and type 2 behaviour is easily and routinely removed, and type 4 
behaviour does not affect the analysis that follows.
It is the presence of these `leptokurtotic' power law tails (also known as 
`heavy tails', or `fat tails') in the PDF of the sampled energies that 
provides the starting point for an analysis of random errors in the estimates 
of expectation values within VMC.

\begin{figure}[t]
\includegraphics{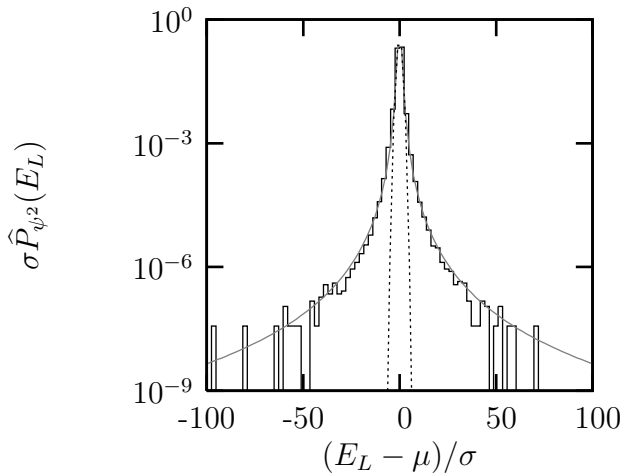}
\caption{\label{fig3}
The seed probability density function estimated by a histogram of $r=10^7$ 
sampled local energies (black).
These are results for an accurate all-electron carbon trial wavefunction, as 
described in the text.
Shown in grey is the model distribution of Eq.~(\ref{eq:4}) that reproduces 
the mean and variance of the samples, and the dotted line is the Normal 
distribution that reproduces the same mean and variance.
}
\end{figure}

Before commencing, it is useful to explicitly show the presence and magnitude 
of the type 3 singularities for a real system, the isolated all-electron 
carbon atom.
A numerical Multi-Configuration-Hartree-Fock calculation was performed to 
generate a multideterminant wavefunction consisting of $48$ Slater 
determinants (corresponding to 7 configuration state functions (CSF)) using 
the ATSP2K code of Fischer \textit{et al.}\cite{fischer07}
Further correlation was introduced via a $83$ parameter Jastrow 
factor\cite{drummond04}, and a $130$ parameter backflow 
transformation\cite{rios06}.
This $219$ parameter trial wavefunction was optimised using a standard 
variance minimisation method\cite{casino06}, resulting in 
$E_{VMC}=-37.83449(7)$ a.u., 
compared with the `exact'\cite{chakravorty93} result of $-37.8450$ a.u.
Of those trial wavefunctions that can practically be constructed and used in 
QMC this may be considered to be accurate, and reproduces $93.3 \%$ of the 
correlation energy at the VMC level.

As discussed above, only type 3 singularities contribute to the asymptotic 
behaviour of the seed distribution.
Figure~\ref{fig3} shows an estimate of the seed PDF, $P_{\psi^2}(E)$, 
constructed by taking $10^7$ standard samples of the local energy, binning 
these into intervals, and normalising\cite{izenman91}.
Also shown is a simple analytic form
\begin{equation}
p(E)=\frac{\sqrt{2}}{\pi} \frac{\widehat{\sigma}^3}{\widehat{\sigma}^4+
    \left(E-\widehat{E}_{tot}\right)^4},
\label{eq:5}
\end{equation}
and a Normal distribution, both with a mean and variance of 
$\widehat{E}_{tot}$ and $\widehat{\sigma}^2$ whose values are obtained from 
the data using the usual unbiased sample estimates.

It is apparent that the seed distribution, $P_{\psi^2}(E)$, is not well 
described by a Normal distribution.
Considering that no fitting procedure is employed (beyond matching the first 
two moments of the model and sample distributions) it is somewhat surprising 
that the simple model distribution is so close to the actual distribution.
This is most clearly demonstrated by comparing the number of sample points 
predicted in a `tail region' defined by 
$(E-\widehat{E}_{tot})>10\widehat{\sigma}=2.25$ a.u.
The numerical data has $2990$ sample points in this region, $p(E)$ predicts 
$3481$ points, and the Normal distribution predicts $1.7\times 10^{-14}$ 
points.

An alternative measure is to assume the asymptote 
\begin{equation}
p_{asym}(E) = \frac{\sqrt{2}}{\pi} \frac{\lambda_3}{\widehat{\sigma}}
       \left( \frac{\widehat{\sigma}}{E-\widehat{E}_{tot}} \right)^{4}
\end{equation}
to be dominant in the `tail region', and to equate the sampled and predicted 
number of outliers.
This estimates the magnitude of the leptokurtotic tails to be $\lambda_3=0.86$ 
(in comparison with $\lambda_3=1$ for the model distribution of 
Eq.~\ref{eq:5}).

Figure \ref{fig3} suggests that the local energy is not well sampled close to 
the nodal surface, where the deviation from the mean is greatest.
Further suspicion that a more detailed analysis is required arises when it is 
noted that third or higher moments do not exist for this seed distribution, 
even though a finite number of samples will provide an estimate of these 
higher moments that converges to infinity as the sample size is increased.

\section{Random error in VMC estimates}
\label{sec:std}
In the previous section no mention of MC methods has been made.
In this section the consequence of choosing the `standard sampling' strategy 
in QMC is investigated.

It has been noted by previous authors that for many calculations the 
distribution of the local energy is clearly not Gaussian, for both VMC and DMC 
calculations\cite{bressanini02,kent99,bianchi88,pandharipande86}.
Section \ref{sec:elsing} shows that this is generally the case.
In previous work it also appears to be implicitly assumed that the form of the 
seed distribution is irrelevant to the application of the CLT to infer 
information on the random error of estimated quantities\cite{foulkes01}.
In what follows, the influence of the leptokurtotic tails on the validity of 
the CLT is examined in detail, and the  distribution of random error in VMC 
estimates is derived.

Numerical evidence for a valid CLT is at best limited, and only weakly 
suggestive.
For most applications of QMC only single estimates are constructed, with an 
estimated random error calculated using the CLT.
Generally no ensemble of estimates is calculated to justify that this error is 
Normal.
The best we can do is observe that for many published results the estimated 
total energies and errors are consistent with exact energies where these are 
known in that they are higher (to within the statistical accuracy suggested by 
the CLT).
This still leaves significant room for non-Gaussian error, especially for 
larger systems and estimates of quantities other than the total energy.

Results for wavefunction optimisation within VMC are more strongly suggestive.
The most stable implementation possible for a stochastic minimisation method 
would provide a Normal random error in the optimised functional.
Instability is commonly observed for many of the available implementations, 
particularly for a large number of particles or where the nodal surface of the 
trial wavefunction is varied\cite{kent99,bressanini02}.
This is consistent with the notion that the CLT may not be valid for these 
implementations.

Possible distributions of error in estimates can be summarised as follows.
The catastrophic case would be for the Law of Large numbers to be invalid, 
providing estimates that do not statistically converge to an expectation as 
$r$ approaches infinity.
Another possibility is that the Central Limit Theorem may not be valid, 
providing estimates that statistically converge, but with a random error that 
is not Normally distributed.
A further possibility is that the CLT may be valid, but that the deviation 
from the Normal distribution for finite $r$ is unknown, so may be significant 
for accessible sample sizes.
A final, ideal case would be for the CLT to be valid, and for the deviation 
from the Normal distribution for finite $r$ to be known, and to be unimportant 
for accessible sample sizes.

The first and last of these are found not to occur, while the other cases do 
(depending on what is being estimated), as a direct consequence of the 
presence of the leptokurtotic tails.

\subsection{Total energy}

As discussed in section \ref{sec:genvmc}, the unbiased estimate of the total 
energy constructed from local energy values at $r$ points sampled from the 
$P_{\psi^2}$ distribution is given by
\begin{equation}
\mathsf{A}_r\left[ E_{tot} \right]  =  
    \frac{1}{r}\sum_{n=1}^{r} \mathsf{E}_n ,
\end{equation}
with $\{ \mathsf{E}_n \}$ the IID random variables 
$E_L(\text{\sffamily\bfseries R})$.
This (rescaled) sum of IID random variables can be analysed using the known 
properties of the PDFs of each $\mathsf{E}_n$ to obtain the PDF of the 
estimate itself.

It is useful to introduce some supplementary random variables in order to keep 
the notation simple.
Defining the mean and variance of $P_{\psi^2}$ as 
$\mathbb{E}\left[ E_L \right]$ and $\sigma^2$ provides the transformation 
\begin{equation}
\mathsf{X}_n=\frac{1}{\sigma}
    \left( \mathsf{E}_n - \mathbb{E}\left[ E_L \right] \right),
\end{equation}
as long as the first two moments exist.
This $\mathsf{X}_n$ has a PDF, $p(x)$, of mean and variance of $0$ and $1$, 
and a symmetric asymptotic behaviour $\propto 1/x^4$.
Two further random variables are $\mathsf{S}_r$, defined as the sum of $r$ 
independent samples taken from $p(x)$, and the normalised version of this sum,
\begin{eqnarray}
\mathsf{Y}_r = \frac{(\mathsf{X}_1+\ldots+\mathsf{X}_r)}{\sqrt{r}}=
    \frac{\mathsf{S}_r}{\sqrt{r}}.
\end{eqnarray}
The transformation from $\mathsf{Y}_r$ to 
$\mathsf{A}_r\left[ E_{tot} \right]$ is 
\begin{equation}
\mathsf{A}_r\left[E_{tot}\right] = 
\frac{\sigma}{\sqrt{r}} \mathsf{Y}_r + \mathbb{E}\left[ E_L \right],
\end{equation}
so that $\mathsf{Y}_r$ is the random error in the estimate of the total energy 
in units of $\sigma/\sqrt{r}$.

The validity of the CLT for these sums of random variables is tested below, 
for the three most common forms of the CLT available.
These are considered in order of increasing generality (in that they are valid 
for progressively larger classes of PDFs) and decreasing knowledge of finite 
sampling effects (in that limits on the deviation from normality for finite 
$r$ are progressively less well defined).

The least general CLT is provided by the existence or not of an Edgeworth 
series expansion\cite{stroock93}.
Provided that all the moments of $p(x)$ exist, and that they satisfy Carleman's 
condition\cite{stroock93}, then the distribution of $\mathsf{Y}_r$ for $r$ 
samples, $P_r\left(y\right)$, can be uniquely defined by the infinite series
\begin{equation}                              
P_r\left(y\right)=\frac{1}{\sqrt{2\pi}} e^{-y^2 /2}
\left(
1 + \frac{f_3(y)}{\sqrt{r}} + \frac{f_{6}(y)}{r} + \ldots
\right) ,
\end{equation}
where each $f_m(y)$ is a finite polynomial in $y$ of order $m$, and with 
coefficients that may be expressed in terms of the first $m$ moments of the 
seed distribution.
If this expansion is valid, $P_r\left(y\right)$ converges to the Normal 
distribution for increasing $r$, and the expansion also provides a definite 
bound on the deviation of the distribution from Normal for finite $r$ - the 
deviation can be estimated if necessary, and scales as the Gaussian function.
For the seed distribution of local energies, $P_{\psi^2}$, the asymptotic 
behaviour ensures that all moments higher than $2^{nd}$ do not exist, hence 
this form of the CLT is invalid.

A more general result is the Berry-Esseen theorem\cite{stroock93}, which 
states that the inequality
\begin{equation}
\left|
\int_{-\infty}^{x} P_r\left( y \right) - \frac{1}{\sqrt{2\pi}} e^{-y^2 /2} dy
\right| \leq
\frac{C}{\sigma^3 \sqrt{r}} \int_{-\infty}^{\infty} |y|^3 p(y) dy,
\end{equation}
is valid provided the $3^{rd}$ absolute moment on the RHS is finite 
($C=0.7655$ is the best value of $C$ available\cite{senatov98}).
This proves that $P_r\left(y\right)$ converges to the Normal distribution for 
increasing $r$, and also provides a bound on the deviation of the distribution 
from Normal for finite $r$.
The asymptotic behaviour of the seed distribution ensures the nonexistence of 
the $3^{rd}$ absolute moment, hence this form of the CLT is invalid for 
$P_{\psi^2}$.

The final candidate is Lindeberg's theorem\cite{stroock93}.
This is the most general form of the CLT, and provides the weakest bound on 
the deviation from Normality for finite $r$.
Provided that 
\begin{equation}
\textrm{Max} \left[ \frac{|\phi(y)|}{1+y^2} \right] < \infty,
\end{equation}
it follows that
\begin{equation}
\lim_{r\rightarrow\infty}
\int_{-\infty}^{x} P_r\left( y \right) \phi(y) dy =
\frac{1}{\sqrt{2\pi}} \int_{-\infty}^{x} \phi(y) e^{-y^2/2}  dy,
\end{equation}
or that in the limit of $r$ approaching infinity the probability of the sum of 
random variables falling in a given interval (given by $\phi(y)=1$) is equal 
to that of the Normal distribution provided by the CLT, provided that the 
$2^{nd}$ moment of $p(x)$ exists.
This provides confidence limits from the sample mean and variance via the CLT 
for large $r$, but two points must be borne in mind.
First, for $\phi(y)$ increasing faster than second order (such as the 
definition of moments higher than $2^{nd}$ order) the expectation is not 
defined, even in the limit of $r$ approaching infinity.
Second, for finite $r$ there is no limit to the magnitude of any deviation 
from Normal, or to how fast these deviations decay with increasing $r$.

These theorems inform us that the random error in the unbiased estimate of the 
total energy obeys the CLT, but no information is available about the 
deviation of the distribution of errors from Normal for finite $r$.
This is unsatisfactory, since only a finite number of samples will ever be 
available.

Using the asymptotic behaviour derived in section \ref{sec:elsing} does allow 
us to extract information about the deviation from Normal that appears in 
$P_r\left( y \right)$.
In what follows this is achieved by using the same strategy as the most 
frequently presented derivation of the CLT\cite{gnedenko68}, but explicitly 
taking into account the leptokurtotic tails.

Denoting the PDF of the sum $\mathsf{S}_r$ as $P_r(s_r)$ (distinct from 
$P_r\left( y \right)$ but related via a change of variables) and viewing this 
sum as a random walk in one dimension leads immediately to the iterative 
convolutions
\begin{equation}
P_r(s_r) = p(x_r) \star P_{r-1}(s_{r-1}),
\end{equation}
starting from $P_1(s_1)=p(x_1)$.
In Fourier space this is simply a product, and defining the Fourier transform 
as
\begin{equation}
p(k)=\int_{-\infty}^{\infty} p(x) e^{-ikx} dx
\end{equation}
immediately gives
\begin{equation}
P_r(k) = e^{r\ln p(k)} ,
\label{eq:6}
\end{equation}
with $P_r(k)$ and $p(k)$ the characteristic functions of $P_r(s_r)$ and $p(x)$ 
respectively.
Equation~(\ref{eq:6}) reduces the problem to that of finding the inverse 
Fourier transform of the $r^{th}$ power of the Fourier transform of the seed 
distribution (with an appropriate transformation of the random variables).

For a PDF to possess a smooth characteristic function (in the sense that all 
derivatives exist at all points), the PDF must decay at least exponentially 
fast as $|x|\rightarrow\infty$\cite{morse53}.
If this were the case, then a Taylor expansion would exist for $\ln p(k)$ that 
is valid for all real $k$.
For the distribution of local energies, the PDF falls to zero algebraically 
slowly which implies the presence of poles in the complex plane for finite 
$|x|$, discontinuities in the Fourier transform at the origin, and no Taylor 
series expansion about $k=0$ for $\ln p(k)$.

The Fourier transform may be performed by contour integration in the complex 
plane, closing the contour in the upper half plane for $\Re[k]< 0$, and the 
lower half plane for $\Re[k]> 0$.
This, in addition to the constraints on the residues and the position of the 
poles that prevent any slower asymptotic behaviour, provides a general series 
expansion
\begin{equation}
\ln p(k) = -\frac{1}{2}k^2 + \frac{\lambda_3}{3\sqrt{2}} \left|k\right|^3 + 
    \eta_3 \left(ik\right)^3 + \mathcal{O}\left(k^4\right).
\end{equation}
All of the coefficients in this expansion are completely unrelated to moments 
of the seed distribution, and for the model distribution shown in 
Fig.~(\ref{fig3}), $\lambda_3=1$ and $\eta_3=0$.
Higher order discontinuities may also be present in this expansion, as 
generally a $|x|^{-q}$ term in the asymptotic behaviour of a function is 
accompanied by a $|k|^{q-1}$ term in its Fourier transform due to the 
properties of bilateral Laplace transforms\cite{morse53}.

This series expansion provides the required expression for $P_r(k)$,
\begin{equation}
P_r(k)= \exp\left[
            -r\frac{1}{2}k^2 +
             r\frac{\lambda_3}{3\sqrt{2}} \left|k\right|^3 + 
             r \eta_3 \left(ik\right)^3 +
            \mathcal{O}\left(k^4\right)
            \right].
\end{equation}
Changing variables to $w=\sqrt{r}k$ and $y=s_r/\sqrt{r}$ and performing the 
inverse Fourier transform gives
\begin{eqnarray}
P_r(y) &=&  \frac{1}{2\pi} \int_{-\infty}^{\infty}
             e^{iwy-w^2/2} \exp\left[
             \frac{\lambda_3}{3\sqrt{2}} \frac{1}{\sqrt{r}} \left| w\right|^3
\right. \nonumber \\ & & \left. +
                   \eta_3                \frac{1}{\sqrt{r}} \left(iw\right)^3 +
             \mathcal{O}\left(\frac{  w^4 }{ r } \right)
             \right] dw,
\end{eqnarray}
where the lowest order terms that are independent of $r$ have been factored 
out.
Expanding the exponential whose argument is a function of $r^{-1}$ as an 
asymptotic series in $r$ gives
\begin{equation}
P_r(y) = \phi_0(y) + 
            \frac{\lambda_3}{3\sqrt{2}} \frac{1}{\sqrt{r}} \chi_3(y) + 
                  \eta_3                \frac{1}{\sqrt{r}} \phi_3(y) + \cdots
\end{equation}
where $\phi_0(y)$ is the standard Normal distribution,
\begin{equation}
\chi_q(y) = \frac{1}{2\pi}\int_{-\infty}^{\infty} 
            \left|w\right|^q e^{iwy-w^2/2} dw
\end{equation}
and
\begin{equation}
\phi_q(y) = \frac{1}{2\pi}\int_{-\infty}^{\infty}
            (i w)^q e^{iwy-w^2/2} dw.
\end{equation}
Higher order terms can be written in the same form, and will have a prefactor 
proportional to $r^{1-q/2}$.
Note that $\chi_q$ and $\phi_q$ are distinct only for odd $q$.

Since $\phi_0(y)$ is a Gaussian function, the CLT is valid, and the PDF may be 
expressed as
\begin{eqnarray}
P_r(y) &=& \frac{1}{\sqrt{2\pi}} \left[
 1 + \frac{\eta_3}{\sqrt{r}} \frac{d^3}{dy^3} 
   + \mathcal{O}\left( \frac{1}{r} \right)
\right] e^{-y^2/2} \nonumber \\ & &
+\left[
     \frac{\lambda_3}{3\pi} \frac{1}{\sqrt{r}} \frac{d^3}{dy^3} 
   + \mathcal{O}\left( \frac{1}{r} \right)
\right] D\left( \frac{y}{\sqrt{2}} \right),
\label{eq:7}
\end{eqnarray}
where $D(x)$ is the Dawson integral\cite{morse53} defined by
\begin{equation}
D\left( x \right) = e^{-x^2}\int_0^{x} e^{t^2} dx,
\end{equation}
and possessing finite derivatives of all orders, and a known asymptotic 
expansion.
Further terms can be included explicitly if required, as higher order 
derivatives of the Gaussian function and Dawson integral.

In a region close to the mean, Eq.~(\ref{eq:7}) may be expanded in the form
\begin{equation}
\lim_{|y| \rightarrow 0} P_r(y) = \left[
\frac{1}{\sqrt{2\pi}} + \frac{1}{\sqrt{r}} h_1\left(y\right) + 
\mathcal{O}\left( \frac{1}{r} \right)
\right] e^{-y^2/2},
\end{equation}
where $h_1$ is an infinite series that converges over a finite region 
surrounding the mean.
This expansion differs from the Edgeworth series in that it does not converge 
for all $y$.

Far from the mean, where the previous series expansion does not converge, the 
asymptotic behaviour takes the form
\begin{equation}
\lim_{|y| \rightarrow \infty} P_r(y) = \left[
  \frac{\sqrt{2}}{\pi}
  \frac{\lambda_3}{\sqrt{r}} \frac{1}{y^4} + 
  \frac{1}{\sqrt{r}}\frac{1}{y^6} h_2\left(\frac{1}{y^2}\right) + 
  \mathcal{O}\left( \frac{1}{r} \right)
\right],
\end{equation}
with $h_2(x)$ an infinite series that converges over a finite region 
surrounding $x=0$.
This form arises because the second sum in Eq.~(\ref{eq:7}) dominates for 
large $y$ (it is obtained from the asymptotic form of the derivative of the 
Dawson integral), and is fundamentally different in character to the Gaussian 
decay that would occur were an Edgeworth series to exists.

The model seed distribution introduced in the discussion of the all-electron 
carbon VMC results of the previous section corresponds to the special case 
$\lambda_3=1$ and $h_1=h_2=0$, and is the simplest form that results in this 
`persistent leptokurtotic' behaviour for the distribution of total energy 
estimates.

These results allow some general observations about the distribution of errors 
in total energy estimates.
As expected, the Normal distribution emerges in the large $r$ limit.
However, for finite $r$ the character of the deviation far from the mean is 
dominated by $E^{-4}$ tails.
The magnitude of these tails, $\lambda_3$ is not expressible in terms of 
moments of the samples, but is required in order to decide whether these 
leptokurtotic tails are statistically significant.

\begin{figure}[t]
\includegraphics{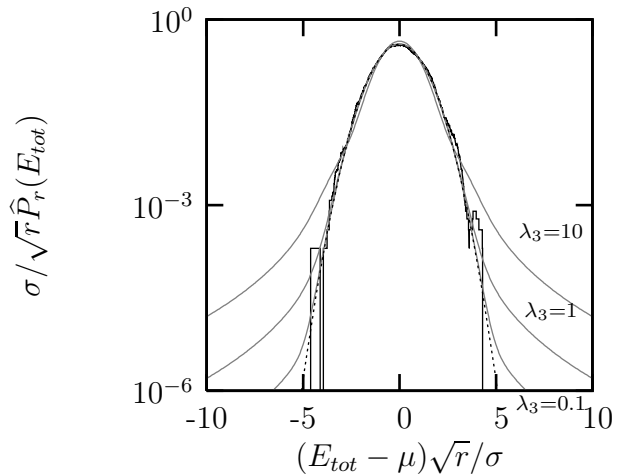}
\caption{\label{fig4}
Probability density function for the random error in the estimated total 
energy.
Results shown are for a kernel estimate of the PDF resulting from $10^4$ 
estimates with $r=10^3$ for each estimate (black).
Grey lines show the predicted distribution, including leptokurtotic tails, for 
different $\lambda_3$ values.
For comparison, the Normal distribution that emerges in the large $r$ limit is 
also shown (dotted line).
}
\end{figure}

Figure~\ref{fig4} shows the distribution of errors, ($P_r$ of Eq.~(\ref{eq:7}) 
truncated to order $1/r^{1/2}$), for a range of $\lambda_3$ values and 
$\eta_3=0$ (a non-zero value would introduce some asymmetry close to the mean).
A non-zero $\lambda_3$ causes a redistribution of probability in an inner 
region where the Gaussian contribution to the density is dominant, with a net 
shift of probability to an outer region where the Gaussian contribution is 
vanishingly small and the leptokurtotic tails dominate.

A useful indicator of the impact of the leptokurtotic tails on confidence 
limits can be extracted as follows.
The deviation from the mean (in units of standard error) at which the 
leptokurtotic tail starts to dominate can be defined as the intersection of 
the dominant parts of the asymptotic and small $y$ expansion of the 
distribution.
This provides the equation
\begin{equation}
y_c^2 = \ln \left( \frac{\pi r}{4\lambda_3^2} \right) + 4\ln y_c^2 ,
\end{equation}
which may be solved numerically, and whose solution depends weakly on 
$r/\lambda_3^2$ due to the logarithmic term.
Specifying extreme values of $r<10^6$ and  $\lambda_3 > 1$ results in 
$y_c < 5.2$.
The value $y_c=4$ is chosen to be representative as it defines the $99.994\%$ 
confidence interval for a Gaussian distribution.
Using this crossover point naturally defines a `Gaussian interval' by $|y|<4$, 
and a `leptokurtotic interval' by  $|y| \geq 4$.
Table~\ref{tab:1} shows the probabilities resulting from a seed distribution 
with varying $r/\lambda_3^2$ values, where a typical value for the carbon atom 
calculations of section~\ref{sec:elsing} is $(r,\lambda_3)=(10^4,1.0)$ or 
$r/\lambda_3^2 = 10^4$.

\begin{table}[t]
\begin{tabular}{rrrr} \hline \hline
                &              & \multicolumn{2}{c}{$\mathrm{Prob}\ (\%)$} \\
$r/\lambda_3^2$ & $\lambda_3$\footnote{corresponding to $r=10^4$}  &
$|\mathsf{y}| \leq 4 $\footnote{Gaussian region} &
$|\mathsf{y}| \geq 4 $\footnote{leptokurtotic region} \\ \hline
$\infty$ &   0.0   &   99.994    &   0.006  \\
$10^6$   &   0.1   &   99.993    &   0.007  \\
$10^4$   &   1.0   &   99.985    &   0.015  \\
$10^2$   &  10.0   &   99.910    &   0.091  \\
$10^1$   &  33.3   &   99.728    &   0.272  \\
$10^0$   & 100.0   &   99.154    &   0.846  \\ \hline \hline
\end{tabular}
\caption{Probabilities of sample total energies in interior and exterior 
regions for a range of values of $r/\lambda_3^2$. $\lambda_3$ values in the 
second column are those corresponding to $r=10^4$.
The range considered is arbitrary, and values that typically arise for 
different systems are unknown.
}
\label{tab:1}
\end{table}

It is apparent that the presence of the leptokurtotic tails could introduce 
significant errors, since the confidence intervals obtained by assuming that 
the error is Normal are not accurate if $r/\lambda_3^2$ is small enough.
For the all-electron carbon atom considered earlier, the Normal interpretation 
appears to be valid provided a confidence of less than $99.98\%$ is required.
For larger $\lambda_3$ the tails become more significant, with outliers 
rapidly becoming more common - the probability of an estimated total energy 
falling in the outlier region increases by two orders of magnitude over the 
range of values shown in the table.

A more direct interpretation of the random error in the total energy can be 
obtained by constructing an estimate of the associated PDF from the numerical 
samples.
A kernel estimate\cite{izenman91} was constructed from $m=10^4$ unbiased total 
energy estimates, each from $r=10^3$ local energy samples using
\begin{equation}
P_r\left(E\right)=\frac{1}{mh} \sum 
   \Theta \left( \frac{E-\mathsf{A}_{r} \left[ E_{tot} \right]}{h} \right),
\end{equation}
where $m$ is the number of estimates, $h$ is the width parameter chosen 
heuristically to provide the clearest plot, and the Kernel function, $\Theta$, 
is chosen to be a centred top-hat function of width $1$.

This (biased) estimate of the PDF is also shown in Figure~\ref{fig4}.
The numerical data provides $1$ sample estimate in the $|y|>4$ region, 
compared with a prediction of $\sim 3$ estimates resulting from the value of 
$\lambda_3=1$ estimated in section~\ref{sec:elsing}.
A Normal distribution (obtained from sample mean and variance and the CLT) 
predicts $0.6$ estimates.
This supports the validity of the CLT confidence limits for these results.

To conclude, estimates of $\lambda_3$ and of the total energy PDF both 
suggest that the leptokurtotic tails are present, but are not statistically 
significant for total energy estimates and the all-electron carbon atom 
considered.
However, it must be borne in mind that the estimated tail magnitude 
($\lambda_3$) has unknown bias, and the range of tail magnitudes for other 
systems is completely unknown.
It seems reasonable to expect a larger, less symmetric system, or a trial 
wavefunction constructed from a finite basis, to provide stronger 
singularities and leptokurtotic tails than the accurate wavefunction 
considered here.
This implies that the degree of validity of a CLT interpretation of confidence 
intervals must be justified for each individual case, a difficult task given 
that no unbiased estimate of $\lambda_3$ is available.

Were leptokurtotic tails to be absent, the evaluation of sample moments would 
be enough to demonstrate that the CLT interpretation was valid, and sample 
moments would provide finite $r$ corrections to the confidence interval.
This is not the case for finite $\lambda_3$ and some (necessarily biased) 
estimate of its value must be obtained from the data.

\subsection{Residual variance}

Following the same approach as for the total energy, the estimate of the 
`variance' of the local energy is considered.
Before analysing the statistics of the standard unbiased estimate for finite 
sample size it is useful to define this quantity in terms of the underlying 
physics of the system, as opposed to the distribution of random samples.
Previous publications\cite{alexander91,kent99,bressanini02} have used distinct 
definitions of the `variance' interchangeably, and inconsistently, especially 
when considering different optimisation and/or sampling strategies.

The residual associated with the Schr\"odinger equation for the system of 
interest and a normalised trial wavefunction, 
$\breve{\psi} = \psi /\left[ \int \psi^2 d\mathbf{R} \right]^{1/2}$, is 
defined as
\begin{equation}
\delta=\left[{\hat H} - E_G\right]\breve{\psi} .
\end{equation}
The `residual variance principle' requires the minimisation of the integral of 
$\delta^2$ over all space with respect to variations in the 
wavefunction\cite{conroy64}.
The parameter $E_G$ may be viewed as a further variational parameter, giving 
the `residual variance' 
\begin{equation}
V_{\delta^2} = 
    \mathbb{E} \left[ \left( E_L - E_{tot} \right)^2 \right]
\end{equation}
where $E_{tot}$ is the expectation value of the total energy of the trial 
wavefunction as defined in the previous section.
This residual variance is zero when $\psi$ is an eigenstate of the 
Hamiltonian, and positive otherwise.

The standard unbiased estimate for this quantity, constructed with `standard 
sampling' and $r$ samples in energy space, is then given by 
\begin{equation}
\mathsf{A}_r\left[ V_{\delta^2} \right] =
  \frac{1}{r-1} \sum_{n=1}^{r}
  \left( \mathsf{E}_n - \mathsf{A}_r\left[ E_{tot} \right] \right)^2.
\label{eq:8}
\end{equation}
In a similar manner to the total energy estimate it is often assumed (whether 
explicitly or implicitly) that the CLT characterises the random error in this 
estimate.

The PDF of this estimate of the residual variance is of interest in its own 
right, as for `standard sampling' it provides the confidence interval for the 
total energy estimate (via the valid CLT assumption for the total energy).
More importantly, the residual variance is often the quantity that is 
minimised when optimising trial wavefunctions, hence the statistics of errors 
in its estimate may well decide the success or failure of an attempt to 
optimise a candidate wavefunction.

In order to express the sum of squares of random variables in Eq.~(\ref{eq:8}) 
as a sum of random variables, $\mathsf{U}_n=\mathsf{X}^2_n-1$ is defined, 
whose PDF can be expressed in terms of the seed distribution $p(x)$ as
\begin{equation}
p_v(u)=
  \frac{1}{2 |u+1|^{1/2}}\left[ p(x=\sqrt{u+1}) + p(x=-\sqrt{u+1}) \right]
\end{equation}
for $u \geq -1$, and $0$ otherwise.
Due to the $x^{-4}$ asymptotic behaviour of the seed distribution, this PDF 
exhibits the asymptotic behaviour
\begin{equation}
\lim_{u\rightarrow\infty} p_v(u) \sim 1/u^{5/2},
\end{equation}
and the second moment of $p_v(u)$ is not defined, hence none of the CLT 
theorems are valid.

From this it follows that the random error in the estimated residual variance 
does not approach a Normal distribution, confidence intervals are not provided 
by the error function, and the sample variance does not provide a measure of 
the random error.
This is the case despite the fact that the sample variance will be finite for 
any number of samples, as it will approach infinity as the number of samples 
is increased.
However, the strong law of large numbers (LLN) is still valid, as $p_v(u)$ 
does possess a finite mean\cite{stroock93}.

A general form of the distribution of the random error is derived in what 
follows, providing a limit theorem that takes the place of the CLT.
The existence of alternative limit theorems (that result in `infinitely 
divisible forms' for the distribution, also known as `Levy skew alpha-stable 
distributions' or `Stable distributions') that are valid for classes of PDF 
functions is well known in statistics,\cite{stroock93,gnedenko68} with the CLT 
and resulting Normal distribution being the most familiar example.

The notation is simplified by defining two supplementary random variables.
A sum of $r$ IID random variables with distribution $p_v(u)$ is denoted 
$\mathsf{S}_r$, and a normalised sum is denoted $\mathsf{V}$, such that
\begin{equation}
\mathsf{V}=\frac{ \mathsf{U}_1 + \ldots + \mathsf{U}_r }{ r^{2/3}} = 
     \frac{\mathsf{S}_r}{r^{2/3}}.
\end{equation}
With these definitions the transformation from $\mathsf{V}$ to 
$\mathsf{A}_r\left[ V_{\delta^2} \right]$ is given by
\begin{equation}
\mathsf{A}_r\left[ V_{\delta^2} \right] = 
   \left(
   \frac{ \mathsf{V} }{ r^{1/3} } + 1
   \right) \sigma^2.
\end{equation}

Following the same approach as for the total energy, the PDF of $\mathsf{S}_r$ 
is given by
\begin{equation}
P_r(s_r) = p_v(u_r) \star P_{r-1}(s_{r-1}),
\end{equation}
and the characteristic functions of $\mathsf{U}$ and $\mathsf{S}_r$ are 
related by
\begin{equation}
P_r(k) = e^{r\ln p_v(k)}.
\end{equation}
In order to continue, a series expansion of the logarithm of $p_v(k)$ is 
required.
For the total energy estimate the analogue of this was obtained by closed 
contour integration in the complex plane, however this is not appropriate for 
$p_v(k)$ due to the presence of fractional powers.
A different route consists of reintroducing the original variable, $x$, into 
the Fourier transform, giving
\begin{equation}
p_v(k)e^{-ik}= \int_{-\infty}^{\infty} p(x) e^{-ikx^2} dx ,
\end{equation}
which may be performed as a bilateral Laplace transform\cite{morse53} to give 
the general series expansion 
\begin{eqnarray}
\ln p_v(k) &=&- \lambda_{3} \frac{4}{3\sqrt{\pi}}
                \left(1 - i\;\textrm{sgn}[k]\right) |k|^{3/2} 
              + \lambda_{4} k^2
 \nonumber \\ & & 
              + \mathcal{O}\left( |k|^{5/2} \right) ,
\end{eqnarray}
where no linear term appears as the mean of $p_v(u)$ is zero (due to the 
offset in the definition of $\mathsf{U}_n$).
Note the discontinuity introduced by a sign function, $\textrm{sgn}[k]$, that 
is equal to $+1$ for positive $k$, $-1$ for negative $k$, and whose definition 
is irrelevant at $k=0$.

This provides the required expression for $P_r(k)$,
\begin{eqnarray}
P_r(k) &=& \exp\left[
            -r \lambda_{3} \frac{4}{3\sqrt{\pi}}
               \left(1 - i\;\textrm{sgn}[k]\right) |k|^{3/2} 
            +r \lambda_{4} k^2 
\right. \nonumber \\ & & \left. \;\;\;\;\;\;\;\;
            +  \mathcal{O}\left( r|k|^{5/2} \right)
            \right].
\end{eqnarray}
Changing variables to $w=r^{2/3}k$ and $v=s_r/r^{2/3}$, and performing the 
inverse Fourier transform results in the PDF of the normalised sum 
$\mathsf{V}$,
\begin{eqnarray}
P_r(v)&=& \frac{1}{2\pi} \int_{-\infty}^{\infty}
             \exp\left[ iwv - \lambda_{3} \frac{4}{3\sqrt{\pi}}
             \left(1 - i\;\textrm{sgn}[w]\right) |w|^{3/2} \right]
\nonumber \\ & & \times
             \exp\left[
             \frac{\lambda_4}{r^{1/3}} w^2 +
             \mathcal{O}\left(\frac{  w^{5/2} }{ r^{2/3} } \right)
             \right] dw.
\end{eqnarray}
The lowest order terms are independent of $r$ due to the normalisation chosen 
for $\mathsf{V}$.
Expanding the second exponential as a power series for large $r$ gives
\begin{equation}
P_r(v) = \chi_{0}(v) + \frac{\lambda_4}{r^{1/3}}\phi_2(v) + \ldots ,
\label{eq:9}
\end{equation}
where
\begin{widetext}
\begin{equation}
\chi_q(v) = \frac{1}{2\pi} \int_{-\infty}^{\infty} |w|^q
\exp\left[
iwv- \lambda_{3} \frac{4}{3\sqrt{\pi}}
\left(1 - i\;\textrm{sgn}[w]\right) |w|^{3/2} \right] dw 
\end{equation}
and
\begin{equation}
\phi_q(v)= \frac{1}{2\pi} \int_{-\infty}^{\infty} (iw)^q
\exp\left[ iwv- \lambda_{3} \frac{4}{3\sqrt{\pi}}
\left(1 - i\;\textrm{sgn}[w]\right) |w|^{3/2} \right] dw ,
\end{equation}
\end{widetext}
and differentiation with respect to $v$ iteratively provides terms of higher 
$q$ from $\chi_0$ and $\phi_0$.
The lowest order term in this expansion provides the distribution of the 
estimate in the large $r$ limit, and is a particular case of the class of 
Stable Distributions\cite{gnedenko68}.

A transformation of the characteristic function to an explicit representation 
of $\chi_0(v)$ is not available in the literature, and is a non-trivial 
integral.
Although a strictly closed form representation is not available, here the 
integral is performed analytically to provide the resulting distribution in a 
concise form employing Bessel functions.
The derivation is given in the appendix, and provides the estimate of the 
residual variance, $\mathsf{A}_r\left[ V_{\delta^2} \right]$, as 
a random variable with a PDF given by 
\begin{widetext}
\begin{equation}
\lim_{r\rightarrow\infty} P_r(x) =
     \frac{\sqrt{3}}{\pi} \frac{1}{2\gamma}
     \left[ \frac{x - \sigma^2}{2\gamma} \right]^2
     \exp{ \left( \left[ \frac{x - \sigma^2}{2\gamma} \right]^3 \right)}
\left[
-\textrm{sgn}    \left[ x - \sigma^2 \right]
 K_{1/3}  \left( \left| \frac{x - \sigma^2}{2\gamma} \right|^3  \right)
+K_{2/3}  \left( \left| \frac{x - \sigma^2}{2\gamma} \right|^3  \right)
\right] ,
\label{eq:10}
\end{equation}
\end{widetext}
where $x$ is a supplementary variable integrated over to obtain probabilities, 
$\sigma^2$ is the variance of the underlying seed distribution of local 
energies, and $\gamma$ is the scale parameter for the distribution defined by
\begin{equation}
\gamma = \left[ \frac{6 \lambda_{3}^2}{\pi r} \right]^{1/3} \sigma^2.
\end{equation}
This distribution of unbiased estimates of the residual variance in `standard 
sampling' takes the place of the Normal distribution that occurs for a valid 
CLT.

The parameter $\lambda_{3}$ is the same as that in the analysis of the total 
energy estimate, and is a measure of the magnitude of the leptokurtotic tails 
in the seed distribution.
The `width' $\gamma$ is not related to the variance of the distribution itself 
- the mean and variance of $P_r(x)$ are $\sigma^2$ and $\infty$ respectively.
Although this width parameter approaches zero for increasing $r$, it does so 
as $r^{-1/3}$ (the analogous width parameter for the CLT decreases as 
$r^{-1/2}$).
The asymptotic behaviour of $P_r(x)$ is given by
\begin{equation}
\lim_{x\rightarrow\infty} P_r(x) =
\frac{1}{2\sqrt{6\pi}}
\frac{1}{2\gamma}
\left( \frac{2\gamma}{x} \right)^{5/2},
\label{eq:11}
\end{equation}
showing that the leptokurtotic behaviour of the PDF for $U=X^2-1$ is preserved.
This is the dominant part of the asymptotic behaviour even for finite $r$, as 
it can easily be shown that the additional terms decay faster than $x^{-5/2}$.

Equation~(\ref{eq:10}) is a general result for the statistics of estimates of 
the residual variance for `standard sampling' in VMC (it is also a general 
result for a sum of IID random variables whose PDF possesses a one sided 
$x^{-5/2}$ asymptote).
General conclusions may be drawn from this distribution.
The most important result is that the CLT does not apply, but the LLN does.
It is apparent that although confidence intervals exist for an estimate of the 
residual variance, they are completely unrelated to a sample variance, and 
confidence intervals obtained using the error function and sample variance are 
unrelated to the distribution of errors even though they could be calculated.

Since no unbiased estimate exists for $\lambda_3$ (or $\gamma$), only the 
biased estimates considered earlier can be used to construct confidence 
intervals.
In addition, closer examination of the form of the distribution reveals that 
the mean may be outside of the confidence interval, since the mode and median 
do not coincide.
Another observation is that, with increasing confidence, a lower bound of the 
confidence interval decreases slowly (slower than the CLT would predict), but 
the upper bound rapidly becomes far larger than that predicted by the CLT.

\begin{figure}[t]
\includegraphics{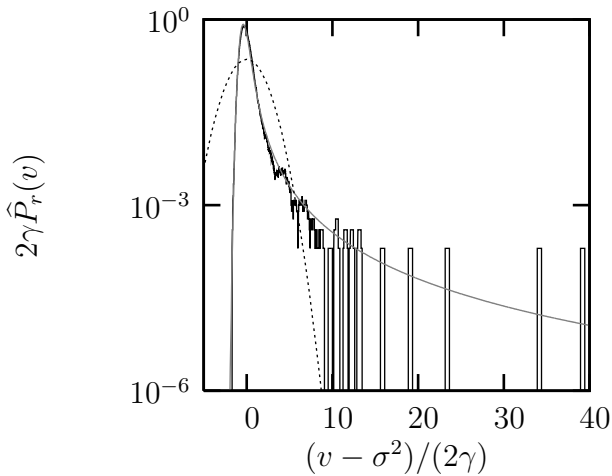}
\caption{\label{fig5} 
Probability density function for the random error in the estimated residual 
variance.
Results shown are for a kernel estimate of the PDF resulting from $10^4$ 
estimates with $r=10^3$ for each estimate (black).
Grey lines show the predicted large $r$ limit (a Stable PDF).
For comparison, a Normal distribution with a mean and variance taken to be the 
sample mean and variance of the data is also shown (dotted line).
}
\end{figure}

Figure~\ref{fig5} shows the general form of the distribution in the limit of 
large $r$, together with a 
kernel estimate of the same distribution constructed from $10^4$ residual 
variance estimates, each from $r=10^3$ local energy samples for the 
all-electron carbon atom considered for total energy estimates.
For comparison, a Normal distribution resulting from blindly applying the CLT 
using the mean and variance of the sampled data is also shown.
It is clear that the distribution of estimated residual variance is far from 
Normal, and it should be remembered that the width of the Normal distribution 
(in units of $2\gamma$) shown in the figure diverges with increasing number of 
residual variance estimates.

Observing that the limiting distribution describes the carbon data well, and 
that $r=10^3$ is a relatively small number of sample points, suggests that the 
large $r$ limit has been reached in this case.
For less accurate trial wavefunctions this may not be the case.
Since the deviation from the large $r$ limit has a magnitude proportional to 
$r^{-1/3}$ this should be justified on a case by case basis.

The significance of the deviation from the Normal distribution may best be 
estimated by considering the predicted number of estimates in the interval 
$(v-\sigma^2)/2\gamma>2$, for $10^4$ estimates.
Incorrectly assuming the validity of the CLT predicts $0.0$ outliers, 
Eq.~(\ref{eq:10}) (with $\lambda_3=1.0$) predicts $\sim 266$ outliers, whereas 
the numerical data provides $198$ estimates in this interval.
Confidence intervals could be defined using Eq.~(\ref{eq:10}) and estimates of 
the parameter $\lambda_3$.
This is not carried out here.
A variety of methods for the estimation of parameters such as $\lambda_3$ do 
exist, but are inherently biased\cite{nolan01}.

It appears that the most important non-Gaussian features of the distribution 
of sample residual variance estimates are that $\gamma \propto r^{-1/3}$, and 
that outliers are likely.

Results for the estimate of both the total energy and the residual variance 
may be summarised in the statement that the `standard sampling' method does 
not sample the $E^{-4}$ tails sufficiently to provide a statistically accurate 
measure of their contribution to estimates.
Were these leptokurtotic tails to be absent, none of the deficiencies 
described above would be present - all moments of the local energy 
distribution would exist, leptokurtotic tails could not occur, and unbiased 
estimates that include finite sample size effects would be readily available.

These results do not invalidate the current use of `standard sampling' for 
total energy or variance estimates, since these estimates still converge to 
the expectation values for increasing $r$.
The difficulty is that estimates of the random error in these quantities are 
not available.
It may be that assuming `$r$ is large enough' provides practical estimates of 
the error in the total energies estimates, but whether this is the case 
depends on more than the sample moments.
Errors in the residual variance estimates are unavoidably not Normal, even in 
the large $r$ limit, and the probability of outliers occurring does not fall 
off exponentially with $r$, but as a power law.

Estimated total energies and residual variance were chosen for consideration 
because of the central role played by these quantities in QMC methods.
In the next section the results of a similar analysis of the 
`standard sampling' estimates for other physical quantities is described, to 
show that the emergence of a non-Normal distribution of errors with power law 
tails is not limited to estimates of the residual variance.

\section{Other Estimates}
\label{sec:otherestimates}
The analysis given in the preceding sections can be applied to general 
estimates in `standard sampling' VMC to obtain the distribution of the 
accompanying random error.
Ideally, it would be hoped that accurate confidence limits would be available 
as a result of the CLT being valid in its strongest form.

In this section estimates of the expectation value of several operators are 
considered, and these take the general form
\begin{equation}
\mathsf{A}_r \left[ X \right] = 
    \frac{1}{r} \sum_{n=1}^{r} x_L( \text{\sffamily\bfseries R}_n ),
\end{equation}
a mean of a local quantity $x_L$.
Singularities in $x_L$ can be classified by location as type 1,2, or 3 in the 
same manner as for the local energy singularities, but the order of the 
singularities is generally different.
The distribution of the estimates themselves are then obtained via the same 
surface integration and generalised central limit theorem approach used for 
the local energy.

\subsection{Kinetic energy and Potential energy}
The most straightforward estimate for the electronic kinetic energy is 
provided by the kinetic part of the local energy,
\begin{equation}
x_L(\text{\sffamily\bfseries R}_n ) = \left[ 
    -\frac{1}{2} \frac{\nabla^2_{\mathbf{R}} \psi}{\psi} 
    \right]_{\text{\sffamily\bfseries R}_n}.
\end{equation}
This possesses type 1 and 2 singularities if the Kato cusp conditions are 
satisfied, and type 3 singularities unless $\nabla^2_{\mathbf{R}} \psi=0$ at 
the nodal surface.
These singularities result in a Normal distribution of estimates in the large 
$r$ limit, with `lopsided' $x^{-4}$ tails in the PDF that decay with 
increasing $r$.
However, the presence of type 1 and 2 singularities is expected to result in 
larger $x^{-4}$ tails in the PDF of the kinetic energy estimate than for the 
total energy estimate.

An alternative estimator for the kinetic energy is provided via Green's 
$1^{st}$ theorem, and takes the form of the sample average of the random 
variable
\begin{equation}
x_L(\text{\sffamily\bfseries R}_n ) = \frac{1}{2} \left[ 
    \sum_i \frac{\nabla_i\psi\cdot \nabla_i\psi}{\psi^2} 
    \right]_{\text{\sffamily\bfseries R}_n},
\end{equation}
where $\nabla_i$ denotes the gradient with respect to the co-ordinate of 
electron $i$.
Type 1 and 2 singularities are not present since the gradient of the 
wavefunction possesses no singularities.
Type 3 singularities arise from the quadratic behaviour of $\psi^2$ about the 
nodal surface, resulting in a positive $x^{-5/2}$ tail in the PDF of the 
sampled random variable and no CLT.
The resulting PDF of kinetic energy estimates is the same one sided Stable PDF 
as for the residual variance estimates, with infinite variance and a 
$x^{-5/2}$ power law tail.

Two potential energy estimates follow naturally from the two kinetic energy 
estimates and the total energy estimate.
One of these possesses type 1 and 2 singularities, and results in a weakly 
valid CLT with strong $x^{-4}$ tails.
The second possesses type 3 singularities only, which result in no valid CLT, 
and the same one sided Stable PDF as the residual variance estimate, with 
infinite variance and a $x^{-5/2}$ power law tail.

\subsection{Non-local Pseudopotentials}
For systems described using non-local pseudopotentials, the local energy 
estimate takes the form
\begin{equation}
x_L(\text{\sffamily\bfseries R}_n ) = \left[ 
    T_L +  V_{ee} + \psi^{-1} \hat{V} \psi 
    \right]_{\text{\sffamily\bfseries R}_n} ,
\end{equation}
where $\hat{V}$ is the sum of one-body non-local operators that make up the 
pseudopotential.
Provided the pseudopotential is not singular these do not possess type 1 
singularities, and type 2 singularities may be prevented using the usual Kato 
cusp conditions.
However, strong type 3 singularities can be expected at the nodal surface, 
resulting in $x^{-4}$ tails in the sample PDF.
Hence, for non-local pseudopotentials, the CLT is expected to be weakly valid, 
with slowly decaying $x^{-4}$ tails that are larger than for the local 
potential case.

\subsection{Mass polarisation and relativistic terms}
Corrections to the total energy due to finite nucleus mass and some 
relativistic effects may be implemented in VMC via perturbation theory, and 
the required estimates are available in the literature\cite{kenny95,vrbik88}.
These generally possess singularities of all three types, and result in 
$x^{-5/2}$ tails in the PDF of the sampled local variable.
As a direct consequence of these tails the CLT is not valid and the large 
sample size limit of the distribution of estimates is not Normal, but a two 
sided variant of the Stable PDF found for the residual variance estimate, that 
is with a finite mean, an infinite variance, and two sided $x^{-5/2}$ power 
law tails.

\subsection{Atomic force estimates}
For estimates of atomic forces the `local Hellmann-Feynman force' is commonly 
taken to possess the form\cite{badinski07}
\begin{equation}
x_L(\text{\sffamily\bfseries R}_n ) = - \left[ 
    \nabla_{\mathbf{X}} \left( \psi^{-1} \hat{V} \psi \right) 
    \right]_{\text{\sffamily\bfseries R}_n} ,
\end{equation}
where $\nabla_{\mathbf{X}}$ is the gradient with respect to the nucleus 
co-ordinate(s), $\mathbf{X}$, evaluated at the nucleus positions of interest, 
and $\hat{V}$ is the sum of one-body potential energy operators due to each 
atomic nucleus in the system.
(Both the operator and the trial wavefunction are functions of the nucleus 
position.)

For the special case where $\hat{V}$ is a local potential the wavefunction 
cancels, and the gradient operator acts on the multiplicative potential only.
For smooth local potentials no singularities arise, and the CLT is valid for 
the resulting estimate.
For a Coulomb potential type 1 singularities arise, and result in estimates 
whose distribution in the large sample size limit is a two sided Stable law of 
finite mean, infinite variance, and with $x^{-5/2}$ power law tails.
For smooth non-local pseudopotentials type 3 singularities arise, and result 
in estimates whose distribution in the large sample size limit is, again, a 
two sided Stable law with $x^{-5/2}$ power law tails.

\subsection{Linearised basis optimisation}
A wavefunction optimisation strategy has recently been 
developed\cite{umrigar07,brown07} that linearises the influence of variational 
parameters on the total energy by constructing a basis set from derivatives of 
the trial wavefunction with respect to parameters of the wavefunction, 
$\alpha_i$.
Applying the total energy variational principle results in a matrix 
diagonalisation problem, with matrix elements defined by integrals that are 
estimated as means of the sample values
\begin{equation}
x_L(\text{\sffamily\bfseries R}_n ) = \left[ 
    \frac{ \psi_i }{\psi} \frac{ \hat{A} \psi_j}{\psi} 
    \right]_{\text{\sffamily\bfseries R}_n},
\end{equation}
with $\psi_i$ the derivative of the trial wavefunction with respect to 
parameters $\alpha_i$, except for $\psi_0=\psi$.
$\hat{A}$ is either the identity or the Hamiltonian operator.

Generally, the linear behaviour of the wavefunction as the nodal surface is 
crossed introduces singularities in the sampled quantity, resulting in 
$x^{-5/2}$ tails in the PDF.
These result in an invalid CLT, and the estimated matrix elements have a PDF 
(in the large sample size limit) of the same form as for the estimate of the 
residual variance - the one sided Stable distribution with infinite variance.
Some exceptions occur for particular matrix elements; for the Hamiltonian 
operator the distribution of the estimate is weakly Normal for $i=0$, and for 
the identity operator the CLT is weakly valid for $i=0$ or $j=0$, and the 
variance is zero for $i=j=0$.

Although this informs us of the distribution of each estimated matrix element, 
it provides no direct information on the correlation between elements, or of 
the distribution of the lowest eigenvalue of the estimated 
matrix\cite{edelman91}.
However, it seems likely that the invalidity of the CLT makes a significant 
contribution to the instabilities that must be carefully controlled for an 
implementation of this optimisation method to be successful.

\section{Conclusion}
\label{sec:conc}

The sampling distribution for a local quantity can be simplified by reducing 
the $3N$-dimensional distribution to the degrees of freedom of the local 
quantity that is sampled, with derivable asymptotic behaviour.
Such an analysis has been applied here to characterise the random error for 
the two most important estimated quantities in variational QMC, the total 
energy and the residual variance.

For estimates of the total energy within the `standard sampling' 
implementation of VMC, the CLT is found to be valid in its weakest form with 
the consequence 
that the influence of finite sample size is not obvious and must be considered 
on a case by case basis.
Outliers have been found to be significantly more likely than suggested by CLT 
confidence limits.
No rigorous bounds exist that provide limits to the deviation from the CLT for 
finite $r$, and consequently confidence intervals based on the CLT may be 
misleading.
However, for the example case of an all-electron isolated carbon atom and an 
accurate trial wavefunction the assumption of large sample size appears to be 
useful.

The variance of the local energy has also been considered in light of the 
primary role played by this and similar quantities in wavefunction 
optimisation procedures.
A statistical variance of the local energy within `standard sampling' is 
equivalent to the residual variance defined in terms of the Hamiltonian and 
trial wavefunction themselves, and the statistics of the estimate of this 
quantity have been investigated.

For estimates of the variance within the `standard sampling' implementation 
the CLT is found to be invalid.
A more general Stable distribution and generalised central limit theorem take 
the place of the Normal distribution and CLT, and this Stable distribution is 
fundamentally different from the Normal distribution.
It possesses tails that decay algebraically, and so outliers are many orders 
of magnitude more likely than suggested by the CLT.
The width scale of this distribution falls as $r^{-1/3}$, significantly slower 
than the $r^{-1/2}$ scaling that would result from a valid CLT.
The distribution is asymmetric, so the mean and mode do not coincide.
Only biased estimates of the parameters of this distribution (other than its 
mean) are available, and confidence intervals based on the CLT are entirely 
invalid.

In order to demonstrate that this is not a statistical issue particular to 
estimating the residual variance, estimates of the expectation values of 
several other operators have also been considered.
For most of these the CLT is found to be invalid, with the same or a similar 
distribution of random error arising as for the residual sampling estimate - 
the Stable distribution with $x^{-5/2}$ asymptotic tails and infinite variance.

Perhaps the most important consequence of these results arises in the context 
of the minimisation of the residual variance and related quantities carried 
out to optimise a trial wavefunction.
Many of the instabilities encountered in different optimisation 
methods\cite{kent99,bressanini02} may be due to the use of estimates that are 
statistically faulty.

By shedding an assumption about the properties of QMC estimates and replacing 
this with a derivation of the true distribution of random errors, it has been 
shown that deviations from the CLT are not trivial and can be expected to have 
a significant influence on the accuracy and reliability of estimated physical 
quantities and optimisation strategies within QMC.
The analysis itself provides a new explicit (but not rigorously closed) 
expression for a particular Stable law PDF, and a general approach to 
assessing the strengths and failures of general sampling strategy/trial 
wavefunction combinations for estimating expectation values of physical 
quantities in QMC.

\begin{acknowledgements}
The author thanks Prof. Richard Needs for helpful discussions, and
financial support was provided by the Engineering and Physical
Sciences Research Council (EPSRC), UK.
\end{acknowledgements}

\appendix*
\section{}
Defining $a^{3/2} = \frac{4\lambda_{3}}{3\sqrt{\pi}}$ gives $\chi_0$ of 
Eq.~(\ref{eq:9}) as
\begin{equation}
\chi_0(v) = \frac{1}{2\pi} \int_{-\infty}^{\infty}
\exp\left[ -  a^{3/2}
\left(1 - i\;\textrm{sgn}[w]\right) |w|^{3/2} \right] e^{iwv} dw .
\end{equation}

Partitioning the integral into the negative and positive ranges gives
\begin{equation}
\chi_0(v) = I_1(v) + I_2(v) ,
\end{equation}
with $I_1$ and $I_2$ integrals taken over $0 \leq w<\infty$ and 
$-\infty < w < 0$, respectively.
Substituting $w=y^2$ results in
\begin{eqnarray}
I_1(v) &=& \frac{1}{2\pi} \int_{0}^{\infty} 
       2y \exp\left[ ivy^2 -  a^{3/2}(1 - i) y^3 \right]dy,
\nonumber \\
\end{eqnarray}
and, for $I_2$, substituting $w=-y^2$ results in
\begin{eqnarray}
I_2(v) &=& \frac{1}{2\pi} \int_{0}^{\infty}  
       2y \exp\left[ -ivy^2 -  a^{3/2}(1 + i) y^3 \right]dy \nonumber \\
       &=& I_1(v)^*.
\end{eqnarray}
These two identities provide 
\begin{eqnarray}
\chi_0(v) &=& I_1(v)+I_1(v)^*           \nonumber \\
          &=& \frac{1}{\pi}  \Re\left[ \int_{0}^{\infty} 
              2y \exp\left[ ivy^2 - a^{3/2}(1 - i) y^3 \right]dy \right]. 
\nonumber \\
\end{eqnarray}
The next step is to obtain the real part of the integral in this expression.
This can be achieved by converting this integral into an ODE for $\chi_0(v)$, 
and then seeking the solutions that are real and normalised.

First define $G_n$ by
\begin{equation}
G_n(v)=\int_{0}^{\infty} 2y^n \exp\left[ ivy^2 -  a^{3/2}(1 - i) y^3\right]dy ,
\end{equation}
so that
\begin{equation}
\chi_0(v) = \frac{1}{\pi}  \Re\left[ G_1(v) \right].
\end{equation}
Equations that relate $G_n$ for different indices may be derived. 
The first of these is obtained by integrating the derivative of the 
exponential function in the integrand to give
\begin{widetext}
\begin{equation}
\int_{0}^{\infty}
  \left( 2ivy - 3a^{3/2}(1 - i) y^2 \right) 
  \exp\left[ ivy^2 -  a^{3/2}(1 - i) y^3\right] dy =
\left. \exp\left[ ivy^2 -  a^{3/2}(1 - i) y^3\right] \right|_{v=0}^{v=\infty}.
\label{eq:a1}
\end{equation}
\end{widetext}
In addition integrating $G_n$ by parts provides the relation
\begin{equation}
(n+1) G_n = -2ivG_{n+2} + 3a^{3/2}(1-i)G_{n+3}.
\label{eq:a2}
\end{equation}
These two expressions provide the equations
\begin{eqnarray}
-1  &=& \;\;\; ivG_1 - \frac{3}{2}a^{3/2}(1-i)G_2 \label{eq:a3} \\
G_1 &=& -ivG_3 - \frac{3}{2}a^{3/2}(1-i)G_4  \\
G_2 &=& -\frac{2}{3}ivG_4 + a^{3/2}(1-i)G_5,
\end{eqnarray}
where the first arises from evaluating the limits in Eq.~(\ref{eq:a1}) 
explicitly and expressing the LHS in terms of $G_1$ and $G_2$ and the 
following two arise from Eq.~(\ref{eq:a2}) for $n=1,2$.

Combining these equations to remove $G_2$ and $G_4$, and noting that 
$\frac{dG_1}{dv}=iG_3(v)$ and $\frac{d^2G_1}{dv^2}=-G_5(v)$ provides
\begin{equation}
9a^3 G_1'' - 2 v^2 G_1' - 5v G_1 = -3i.
\end{equation}

Making the substitutions
\begin{equation}
G_1(v)=v^2 e^{\left( \frac{v}{3a} \right)^3} g(v)
\end{equation}
and
\begin{equation}
x=\left( \frac{v}{3a} \right)^3,
\end{equation}
further simplifies this ODE, and results in the inhomogeneous ODE
\begin{equation}
 x^2g'' +  2 x g' - \left(x^2+x-\frac{2}{9}\right)g = -\frac{1}{27a^3}i e^{-x}.
\end{equation}
Only the real solutions of this equations are required, hence only the 
homogeneous ODE
\begin{equation}
 x^2g'' +  2 x g' - \left(x^2+x-\frac{2}{9}\right)g = 0
\end{equation}
need be considered.
The required solution is finite for $x\rightarrow \pm \infty$ and continuous 
at $x=0$, and is a sum of two modified Bessel functions of the second kind,
\begin{equation}
g(x) = A\left[
-\textrm{sgn}(x) K_{1/3}\left( \left| x \right| \right) +
        K_{2/3}\left( \left| x \right| \right)
\right],
\end{equation}
with $A$ an undefined constant.

Requiring Eq.~(\ref{eq:a3}) to be true for $v=0$ provides $A$, and 
transforming back to $v$ provides the final result
\begin{widetext}
\begin{equation}
\chi_0(v) =
\frac{\sqrt{3}}{\pi}
\frac{v^2}{(3a)^3}
e^{ (v/(3a))^3 } \left[
-\textrm{sgn}(v) K_{1/3}\left( \left| \frac{v}{3a} \right|^3 \right) +
          K_{2/3}\left( \left| \frac{v}{3a} \right|^3 \right)
\right].
\label{eq:a4}
\end{equation}
\end{widetext}
The transformation between $v$ and a more general variable is described in the 
main text.

This provides an explicit form for the PDF of the Stable distribution 
${\bf S} \left(3/2,-1,\gamma,\delta;1\right)$ (using the notation of 
Nolan\cite{nolan01}) - Eq.~(\ref{eq:a4}) is for $(\gamma,\delta)=(a,0)$ and 
the general form is trivially related to this by rescaling and translation.


\end{document}